\documentclass[runningheads]{llncs}
\usepackage[T1]{fontenc}
\usepackage{amsmath}
\usepackage{acro}
\usepackage{subcaption}
\usepackage{graphicx}

\DeclareAcronym{AM}{
  short = AM,
  long = Additive Manufacturing
}

\DeclareAcronym{API}{
  short = API,
  long = Application Programming Interface
}

\DeclareAcronym{DDL}{
  short = DDL,
  long = Data Definition Language
}

\DeclareAcronym{FBNet}{
  short = FBNet,
  long = Facebook-Berkeley-Net
}

\DeclareAcronym{MCUNet}{
  short = MCUNet,
  long = Microcontroller Unit Network
}

\DeclareAcronym{MLP}{
  short = MLP,
  long = Multilayer Perceptron
}

\DeclareAcronym{CV}{
  short = CV,
  long = Computer Vision
}

\DeclareAcronym{NAS}{
  short = NAS,
  long = Neural Architecture Search
}

\DeclareAcronym{NN}{
  short = NN,
  long = Neural Network
}

\DeclareAcronym{FPGA}{
  short = FPGA,
  long = Field Programmable Gate Array
}

\DeclareAcronym{FLOP}{
  short = FLOP,
  long = Floating Point Operation
}

\DeclareAcronym{LPBF}{
    short = LPBF,
    long = Laser Powder Bed Fusion 
}

\DeclareAcronym{GPU}{
  short = GPU,
  long = Graphics Processing Unit
}

\DeclareAcronym{CPU}{
  short = CPU,
  long = Central Processing Unit
}

\DeclareAcronym{CNN}{
  short = CNN,
  long = Convolutional Neural Network
}

\DeclareAcronym{ML}{
  short = ML,
  long = Machine Learning
}

\DeclareAcronym{NIC}{
  short = NIC,
  long = Network Interface Card
}

\DeclareAcronym{PCIe}{
  short = PCIe,
  long = Peripheral Component Interconnect Express
}

\DeclareAcronym{HPC}{
  short = HPC,
  long = High-Performance Computing
}

\DeclareAcronym{HPO}{
  short = HPO,
  long = Hyperparameter Optimization
}

\DeclareAcronym{AI}{
  short = AI,
  long = Artificial Intelligence
}

\DeclareAcronym{RSME}{
  short = RSME,
  long = Root Mean Square Error
}

\DeclareAcronym{EA}{
  short = EA,
  long = Evolutionary Algorithms 
}

\usepackage{hyperref}
\usepackage{color}

\usepackage[capitalize]{cleveref}
\begin{document}
\title{Optimizing edge AI models on HPC systems with the 
edge in the loop}
\author{Marcel Aach\inst{1}\orcidID{0000-0002-7861-0672}
\and
Cyril Blanc\inst{2}\orcidID{0000-0003-3271-2398} 
\and
Andreas Lintermann\inst{1}\orcidID{0000-0003-3321-6599}
\and
Kurt De~Grave\inst{2}\orcidID{0000-0001-9116-6986}}
\authorrunning{M. Aach, C. Blanc, A. Lintermann, and K. De Grave}
\institute{Jülich Supercomputing Centre, Jülich, Germany
\email{m.aach@fz-juelich.de, a.lintermann@fz-juelich.de}\\
\and
ProductionS core lab, Flanders Make, Lommel/Leuven, Belgium
\\
\email{cyril.blanc@flandersmake.be, kurt.degrave@flandersmake.be}}%
\maketitle

\begin{abstract}
\ac{AI} and \ac{ML} models deployed on edge devices, e.g., for quality control in \ac{AM}, are frequently small in size. Such models usually have to deliver highly accurate results within a short time frame. Methods that are commonly employed in literature start out with larger trained models and try to reduce their memory and latency footprint by structural pruning, knowledge distillation, or quantization. It is, however, also possible to leverage hardware-aware \ac{NAS}, an approach that seeks to systematically explore the architecture space to find optimized configurations. In this study, a hardware-aware \ac{NAS} workflow is introduced that couples an edge device located in Belgium with a powerful \ac{HPC} system in Germany, to train possible architecture candidates as fast as possible while performing real-time latency measurements on the target hardware. The approach is verified on a use case in the \ac{AM} domain, based on the open RAISE-LPBF dataset, achieving $\approx 8.8$ times faster inference speed while simultaneously enhancing model quality by a factor of $\approx 1.35$, compared to a human-designed baseline.

\end{abstract}

\section{Introduction}
\acresetall
Deploying \ac{ML} models on edge devices presents unique challenges, as these systems must deliver high accuracy while operating under strict memory and latency constraints. Edge \ac{AI} is widely used in applications requiring real-time decision-making. This includes industrial automation and process monitoring, where traditionally post-training optimization approaches like pruning and quantization are applied. \\

An alternative is hardware-aware \ac{NAS}, which systematically explores model architectures to find a best-suited configuration for a given hardware platform. This study introduces a \ac{NAS} workflow that pairs an edge device located in Belgium with a \ac{HPC} system in Germany. This setup accelerates model training while simultaneously optimizing inference speed on the target hardware, ensuring a practical, improved, and efficient deployment. The corresponding code of the HPC2Edge workflow is available open-source on GitHub\footnote{HPC2Edge GitHub: \url{https://github.com/Flanders-Make-vzw/HPC2edge}}. \\

The approach is validated with a \ac{LPBF} application, an industrial \ac{AM} process that fabricates metal parts. Real-time anomaly detection is essential for preventing defects and reducing waste. The method uses a 20 kHz high-speed camera and a \ac{NN}-based video regression model to predict laser parameters. Deviations of the predicted laser parameters from ground-truth laser parameters indicate process anomalies~\cite{booth22}. For training and evaluation of the method, the RAISE-LPBF-Laser dataset (v1.1)\footnote{RAISE-LPBF-Laser dataset: \url{https://www.makebench.eu}}~\cite{blanc2023reference}, consisting of high-speed camera frames paired with various laser parameters, is used. Optimizing inference speed without sacrificing accuracy is key. Deploying the \ac{NAS}-optimized model presented in this study on an edge device ensures seamless vision integration on any LPBF machine, while improving efficiency and reliability.

The paper is structured as follows: Sec.~\ref{sec:related_work} summarizes the related work on hardware-aware \ac{NAS}, Sec.~\ref{sec:workflow_design} describes the developed workflow in detail, and Sec.~\ref{sec:empirical_results} presents the empirical results. Finally, Sec.~\ref{sec:summary} provides a summary and a conclusion. 

\section{Related Work on Hardware-Aware \ac{NAS}}\label{sec:related_work}
To leverage high-performing \ac{ML} and \ac{AI} models in a practical setting on resource-limited edge devices, two approaches exist. On the one hand, an already optimized model is compressed to fit on the hardware, e.g., by quantization or structural pruning. On the other hand, hardware-aware \ac{NAS} seeks to find the optimal building blocks for a model and then constructs its architecture from scratch. While in regular \ac{NAS} the objective is to find the best performing \ac{NN} architectures in terms of accuracy, hardware-aware \ac{NAS} is inherently multi-objective as not only the accuracy of a model but also factors such as the model size and inference speed are of high relevance. Several methods for performing hardware-aware \ac{NAS} for different types of edge devices have already been introduced in the literature and are summarized in the following, based on a general overview of the field in~\cite{benmeziane2021comprehensivesurveyhardwareawareneural}. \acp{FBNet}~\cite{fbnets}, a family of convolutional architectures for use on mobile devices were discovered using a differential \ac{NAS} approach and outperformed human crafted architectures (such as MobileNets~\cite{howard2017mobilenetsefficientconvolutionalneural}) at the time in terms of speed and accuracy. FNAS~\cite{fnas} leverages hardware-aware \ac{NAS} for creating \ac{NN} architectures that meet the specifications of \acp{FPGA}. It uses a multi-objective reinforcement learning \ac{NAS} approach~\cite{zoph2017neural}, where the latency of an architecture candidate is estimated and only verified after the \ac{NAS} run on the target \ac{FPGA}. The \ac{MCUNet} in~\cite{mcunet} focuses on microcontroller units, which feature even smaller memory than mobile phones. It also leverages a two stage process, where the \ac{NAS} search space is first refined, such that all possible candidates fit the resource constraints of the edge device. Then, the \ac{NAS} for the architecture with the best accuracy is launched. The memory footprint and the \ac{FLOP} performance are calculated not on the edge device. From an optimization technique point of view, also \ac{EA} are a strong choice~\cite{eiben2003introduction}. In \ac{EA}, an initial population is sampled randomly. Subsequent generations are iteratively obtained from the previous one through selection (biased for fitness), mutations, and usually also crossover, i.e., sex. Measurement of fitness, which requires fully training the \acp{NN} candidates is here the expensive step. \\

A critical aspect of hardware-aware \ac{NAS} is accurately measuring hardware costs. While the number of parameters and \acp{FLOP} required for the inference of an architecture candidate can be easily estimated, it has been shown that other quantities of interest, such as the inference time, cannot be reliably derived from these. This is, for instance, the case on different types of edge devices and especially relevant when \acp{GPU} are used~\cite{li2021hwnasbench}. For execution latency, real-world measurement has shown to be the most accurate technique. This may, however, increase the runtime of the \ac{NAS}, as each network candidate needs to be transferred to the edge device, perform the measurement, and return the results. Therefore, many works rely on learning a surrogate model, use a look-up table or heuristics, to predict the latency on the target hardware. Even \ac{ML}-based prediction models result in an error that is off by a factor of up to 3.8, compared to the actual latency~\cite{benmeziane2021comprehensivesurveyhardwareawareneural}. 
Other important hardware cost measurements include energy consumption and memory footprint. While benchmarks exist that collect a large number of edge measurements on modern devices, they are often limited in scope. For \ac{CV} workloads these are mainly focused on \acp{CNN}~\cite{li2021hwnasbench}, while the ones that focus on Transformer-based models tend to emphasize large language models.~\cite{sukthanker2024hwgptbenchhardwareawarearchitecturebenchmark}.

\section{Design and Implementation}\label{sec:workflow_design}
This section introduces the database schema in Sec.~\ref{sec:database_schema}, the \ac{AI} model along with its architectural and optimizer-related hyperparameters in Sec.~\ref{sec:model_description}, and the setup of main HPC2Edge workflow in Sec.~\ref{sec:workflow_setup}. \\

To enable a wide roll-out of monitoring of 3D printers, it is highly preferable to run the inference on embedded hardware near the printer rather than remotely and expensively on a power-hungry machine. Therefore, the community ultimately faces a multi-objective optimization problem: finding a model that is as accurate as possible and at the same time sufficiently fast and small for inferencing on embedded hardware.

\begin{figure}
    \centering
    \includegraphics[width=0.3\linewidth]{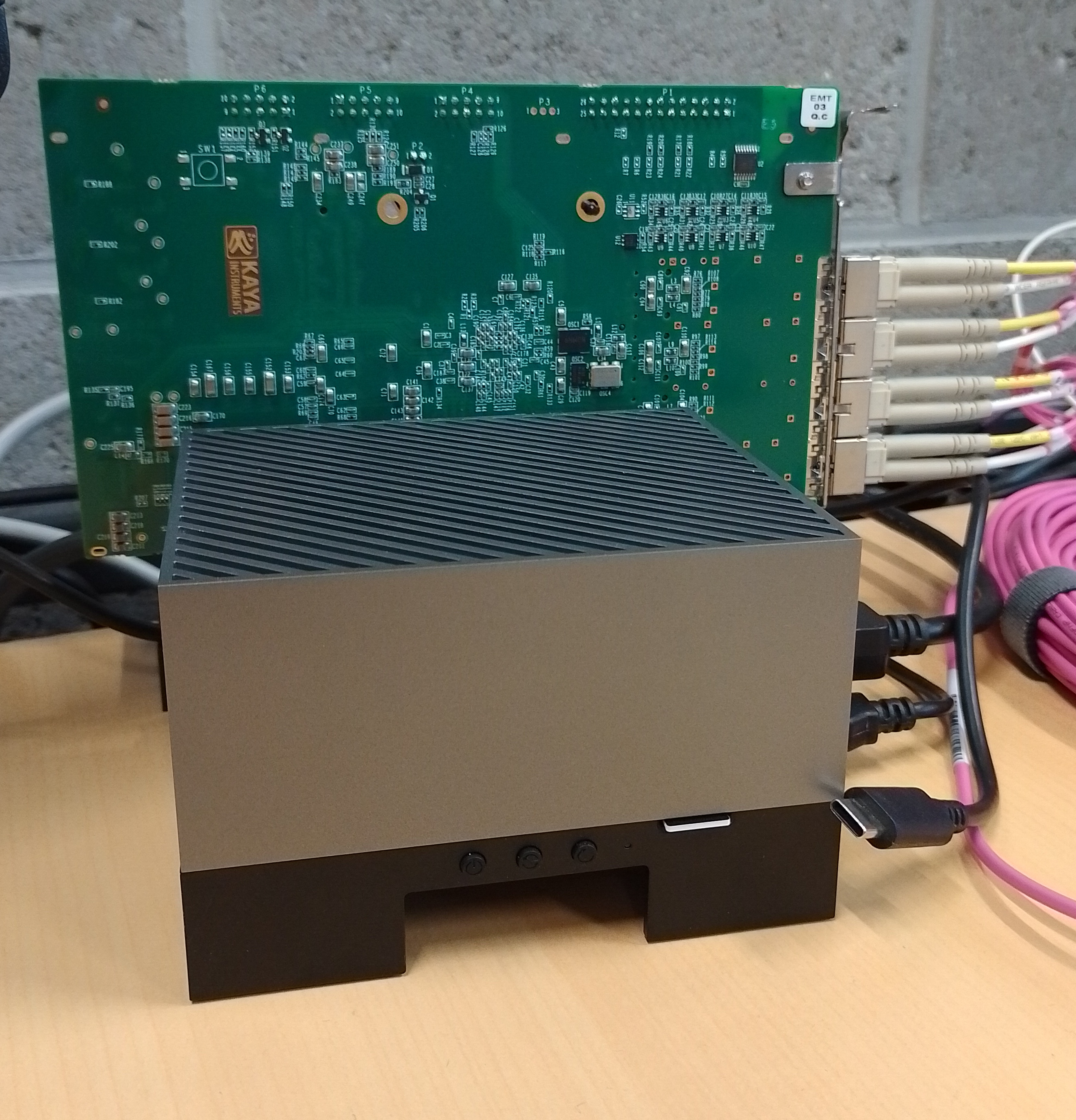}
    \caption{The edge device, an Nvidia AGX Orin (front), with a frame grabber PCIe card (green) for interfacing with high-speed cameras over fiber.}
    \label{fig:edge-device}
\end{figure}

The embedded system of choice is an NVIDIA Jetson AGX Orin™ system, see~\cref{fig:edge-device}. This system is fairly powerful and expensive ($\sim$2,000 EUR as of 03/2025) for an embedded device. The cost is, however, reasonable compared to the much more expensive metal printer and camera. The Jetson has an integrated 10~Gbps Ethernet \ac{NIC}, a \ac{PCIe} slot for hosting a frame grabber, and it can emulate smaller and cheaper devices of the same series. System and \ac{GPU} memory are unified on the board. On the full AGX Orin, memory is not a constraining factor for storing the default \ac{NN} architecture of the baseline model presented in this study. However, the speed of inference remains a constraint, as the system must be able to process the entire surface to catch all faults. The latency of this prediction should be low, i.e., feedback to the controller arrives within a few scanlines to avoid more damage and allow recovery of the fault. Ideally, this processing time should be not much longer than the time it takes to print a scanline.

The two objectives are therefore (i) inference speed and (ii) the \ac{RSME} of the predictions. The inference speed for a model architecture is influenced by many factors, such as the number of parameters. It can be roughly estimated/interpolated (see Sec.~\ref{sec:related_work}), but only an execution on the device itself can reveal the true inference performance. Therefore, the inference speed of all model variants considered in this study are directly measured on the embedded device. \\

A central relational database (in PostgreSQL) has been set up to allow bi-directional communication between the embedded device located in Belgium at Flanders Make and the \ac{HPC} cluster located at the Jülich Supercomputing Centre, Forschungszentrum Jülich, in Germany. The schema is shown in~\cref{fig:datamodel-hpc2edge}.
The optimizer consults the embedded device as soon as it conceives a new candidate \ac{NN} architecture (hyperparameter setting), posting the architecture details to the database. The embedded device continuously polls the database for unmeasured architectures, compiles and optimizes the architecture with the NVIDIA TensorRT library\footnote{NVIDIA TensorRT: \url{https://developer.nvidia.com/tensorrt-getting-started}}, and (after warmup) runs a few inference steps to measure steady-state latency and throughput at several batch sizes. It subsequently reports its results back to the database. \\

The current setup lacks load balancing for multiple embedded devices of the same type, which is desirable for extremely large-scale optimizations, for efficient parallel operation also on the embedded side, as well as to achieve a degree of fault tolerance. At this point, further optimizations, such as the introduction of surrogate models, also become relevant.

\subsection{Database Schema}\label{sec:database_schema}
\begin{figure}[tbh]
    \centering
    \includegraphics[width=\linewidth]{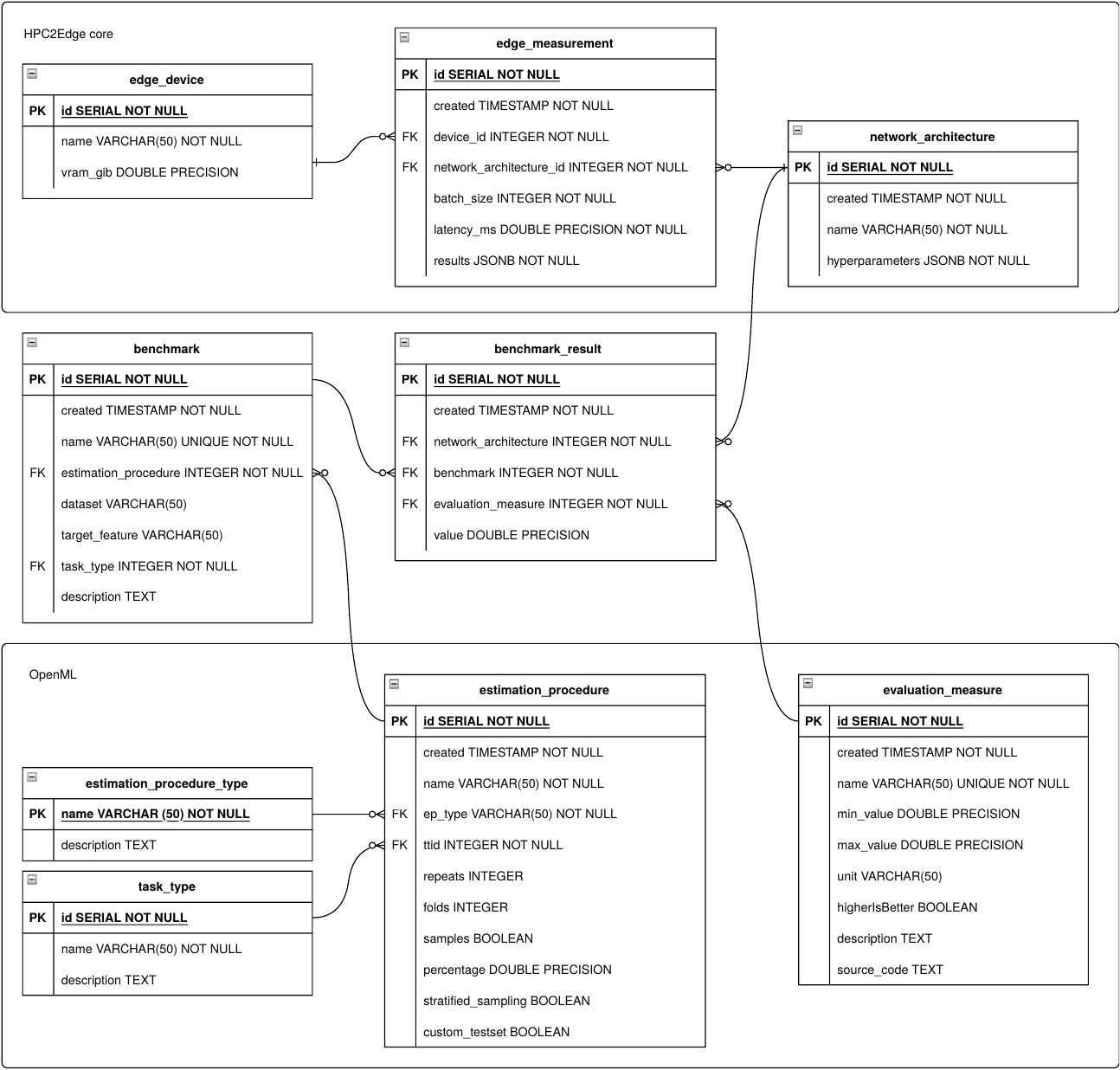}
    \caption{Relational database schema for connecting the HPC-based HPO with an embedded device for inference measurements.}
    \label{fig:datamodel-hpc2edge}
\end{figure}

The HPC2Edge database schema, shown in Fig.~\ref{fig:datamodel-hpc2edge}, consists of an essential part that supports basic communication between the \ac{HPC} and edge systems (labeled `HPC2Edge core' in the figure), and accessory tables to optionally store the full exploration of the optimizer.
In terms of core schema, the edge device is \texttt{GRANT}ed \texttt{INSERT} permission only into the \texttt{edge\_measurement} table.  The \ac{HPO} algorithm gets an account that can \texttt{INSERT} into the (neural) \texttt{network\_architecture} and \texttt{benchmark\_result} tables. All accounts can \texttt{SELECT} from all tables. The \texttt{JSONB} columns allow noSQL-equivalent freedom to evolve the system without changing the main schema, but can still be indexed and efficiently queried when needed. \\

The extended schema is designed to be compatible with OpenML\footnote{OpenML: \url{https://openml.org}}~\cite{OpenML2013}, which is an open platform for sharing datasets, algorithms, and experiments. OpenML has similarities to the more recent Hugging Face\footnote{Hugging Face: \url{https://huggingface.co}} platform, it is, however, more geared towards classical \ac{ML} using tabular datasets.  It offers \acp{API} and supports experiment logging from several popular \ac{ML} toolkits.
The present work uses the publicly available code as of 08/2024\footnote{OpenML 08/2024: \url{https://github.com/openml/OpenML/tree/develop/data/sql}}.
Note that no full direct compatibility is achieved and OpenML has announced a full backend code rewrite, i.e., their future schema might be structured substantially different. \\

OpenML uses an extremely flexible, untyped schema.  Here,  some untyped, string-serialized fields were specialized to double precision, and the table \texttt{math\_function} to \texttt{estimation\_procedure}. Only reference records relevant for regression are stored, without loss of generality. The tables \texttt{benchmark} and \texttt{benchmark\_result} correspond conceptually to tasks and runs in OpenML. It should be noted that the corresponding \ac{DDL} was not available publicly to ensure some level of compatibility. Future work may consider running a full OpenML server for experiment logging --- or some other logging method like MLflow~\footnote{MLflow: \url{https://mlflow.org/}} or ClearML~\footnote{ClearML: \url{https://clear.ml/}} --- extended with only the HPC2Edge core schema.

\subsection{AI Model}\label{sec:model_description}
The model used to predict the laser parameters and to produce the following results (see Sec.~\ref{sec:empirical_results}) is a Video Swin Transformer~\cite{liu2021video} with a modified fully connected end layer for power and speed regression. The data pre-processing is the same as described in~\cite{blanc2023reference}. The RAISE-LPBF-Laser dataset (v1.1)~\cite{blanc2023reference}, consisting of high-speed camera frames paired with various laser parameters, is used. Training and validation focus on a single object (C027) with an 80-20 split, while object C028 is used for testing. The proposed \ac{NAS} framework directly optimizes performance for edge deployment, balancing both speed and accuracy, while leveraging \ac{HPC} for acceleration. This work provides an efficient solution for real-time \ac{AI}-driven industrial applications. \\

The input to the model consists of a window of $16$ consecutive frames that are randomly sampled from each scanline and then normalized and resized to a model input shape of $(256,256)$ pixels.
The output of the model corresponds to the ground-truth values, consisting of a pair of setpoints for laser dot speed and power for each scanline, which are normalized by dividing by their nominal values of 900$mm/s$ and 215$W$, respectively. The choice of using a Video Swin model for prediction is motivated by the fact that it is the best performing attention-based model from~\cite{liu2021video} and that its architecture can be easily modified. Specifically, the model's hyperparameters are well-designed to minimize conflicts and interdependencies, reducing the likelihood of parameterization issues during the optimization run. \\

The hyperparameters of the model that are optimized during the \ac{NAS} run are listed in Tab.~\ref{tab:hyperparams}. The search space includes various Transformer-specific architectural parameters, i.e., the video patch size, which controls the temporal and spatial granularity of the input, the embedded dimensions influencing the dimensions of the tokens, the depths of each model stage, the number of attention heads, the window size of the self attention, and the ratio of feed-forward \ac{MLP} layers between attention blocks. The classical optimizer-related parameters are the base learning rate of the Adam optimizer~\cite{kingma2017adammethodstochasticoptimization} and the scheduler-specific step-size and learning rate decay factor. The search space is chosen to be high-dimensional to allow for an extensive exploration of model size and model quality. 

\begin{table}
\caption{Hyperparameter search space, consisting of architectural and optimizer-related hyperparameters of the Video Swin Transformer model. }\label{tab:hyperparams}
\begin{tabular}{l|p{0.37\linewidth}|l|l}
\textbf{Name} & \textbf{Description} & \textbf{Default} & \textbf{Sampling Range} \\
\hline
Patch size & Video patch size for transformer tokenization & [2,4,4] & [2, 4] each \\
Embedded dimensions & Number of linear projection output channels & 96 & [24, 48]\\
Depths & Depths of each Video Swin Transformer stage & [2,2,6,2] & [1, 2, 4] each \\ 
Heads number & Number of attention heads of each stage & [3,6,12,24] & [3,6,12,24] each\\
\ac{MLP} ratio & Ratio of \ac{MLP} hidden dim. to embedding dim. & 4 & [1, 2, 3, 4]\\
Learning rate & Controls how much to adjust model weights during training & $1 e^{-4}$ & log{[}1e-5, 1{]} \\
Learning rate step size & Interval of learning rate adjustment & 10 & [10, 20, 40] \\
Learning rate $\gamma$ & Learning rate decay factor & 0.5 & (0.1, 0.9)\\
\end{tabular}
\end{table}

\subsection{Workflow Setup}\label{sec:workflow_setup}
\begin{figure}[h]
    \centering
    \includegraphics[width=0.8\linewidth]{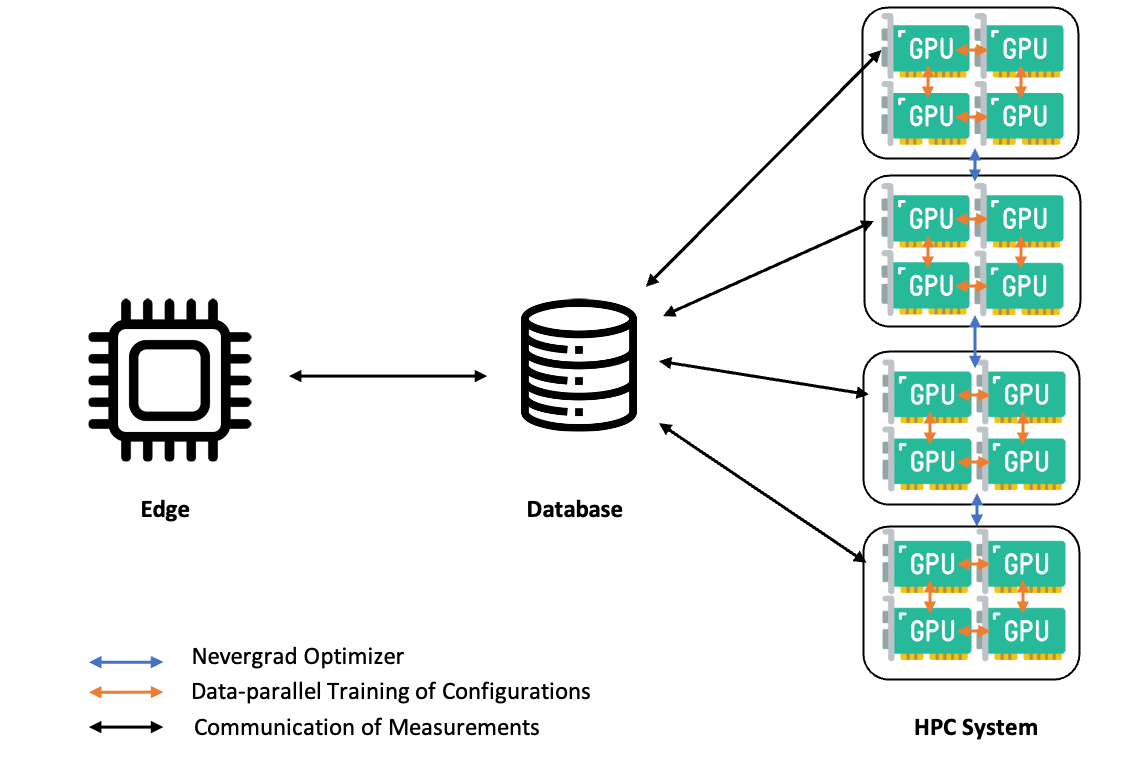}
    \caption{Orchestration of the Hardware-aware NAS search, with communication between the \ac{HPC} system, located at the Jülich Supercomputing Centre, Forschungszentrum Jülich, in Germany, and the edge device, located at Flanders Make in Belgium.}
    \label{fig:hpc2edge_workflow}
\end{figure}

The setup of the HPC2Edge workflow is shown in Fig.~\ref{fig:hpc2edge_workflow}. The training of the different models is performed on the Extreme-Scale Booster partition of the DEEP-EST \ac{HPC} machine~\cite{deep_est} at the Jülich Supercomputing Centre, Forschungszentrum Jülich, in Germany. It features a total of 75 nodes, each one equipped with one NVIDIA V100 GPU and an Intel Xeon 4215 \ac{CPU} with 8 cores and a base frequency of 2.5 GHz. To achieve results in a reasonable amount of time, the training of the different \acp{NN} is performed in data-parallel fashion with the PyTorch-DDP library\footnote{PyTorch-DDP: \url{https://pytorch.org/docs/stable/notes/ddp.html}}. Orchestration of the \ac{HPO} runs is handled by the Ray Tune framework\footnote{Ray Tune: \url{https://www.ray.io/}}. The optimization process leverages the Nevergrad library\footnote{Nevergrad: \url{https://facebookresearch.github.io/nevergrad/}}, a gradient-free optimization tool. Nevergrad performs evolutionary optimization in settings where the computation of gradients is hard or impossible. It is, therefore, a suitable solution for black box optimization problems such as \ac{HPO} and \ac{NAS}. It features a variety of optimization methods, that can be selected based on the search space and available computing budget. For the present work, the $(1+1)$ \ac{EA} was chosen. The algorithm starts with an initial parent population and then creates one offspring for each parent via mutation. It subsequently evaluates the fitness of both the parent and the offspring. In case the offspring achieves a better fitness value than the parent, it replaces the parent in the subsequent generation. For the present experiments, the population size is fixed at $8$, while the total number of evaluations is varied from $16$ to $64$. \\

The edge device periodically queries the database for new, unevaluated entries that match its configuration, i.e., for supported edge device types. Upon finding a relevant entry, it loads the model parameters and performs ten inference runs to compute an average timing. This process is repeated for each configured batch size (1, 2, 4, and 8 in this case). The measured inference times, along with other measurements not exploited in this method, e.g., memory usage, \ac{CPU} usage, or \ac{GPU} usage, are entered in the database to be leveraged by the optimizer.
Once a hyperparameter candidate is chosen, four \acp{GPU} are allocated to its training. With a population size of $8$, this results in $32$ \acp{GPU} being used at the same time. Before launching the training, the head \ac{GPU} submits the architectural details to the database for latency measurement on the edge device. After training for two epochs, the head node reads back this runtime measurement and combines it with the achieved validation loss. Submitting the architecture to the edge device before training the model and inquiring about the runtime measurement only after the model is trained hides the latency of communication between the \ac{HPC} system and edge device. 
\begin{align}\label{eq:solution_combination}
    score_{val} = loss_{val} \cdot 1000 + time_{inference}
\end{align}

A weighted validation score value, based on validation loss and inference time in milliseconds (see Eq.~\ref{eq:solution_combination}) is then reported back to the optimizer and minimized. The best performing model is chosen according to the lowest score achieved. This model is then evaluated on the unseen test dataset, where also a test score is computed in a similar way. 

\section{Empirical Results}\label{sec:empirical_results}
The empirical results of running the hybrid workflow are shown in Tab.~\ref{tab:results_table}. It compares the default hyperparameter configuration, which was chosen by an expert (baseline) based on experience and several experiments, against running hardware-aware \ac{NAS} with an increasing number of samples $n = \{16, 32, 64\}$. The training times of a single configuration range between 1~--~3 hours, while the whole \ac{HPO} run on the \ac{HPC} system took between 8~--~19 hours, stretching the maximum allowed job time of 20 hours on the \ac{HPC} system. The evaluation metrics include the validation loss $l_v$, the inferences time $t$ and the test loss $l_t$ of the best configuration, which is chosen according to the lowest validation score metric $s_v$, see Eq.~\eqref{eq:solution_combination}. The most significant reduction in comparison to the baseline can be observed for the inference time metric. Using just $16$ samples decreases the inference time by a factor of $\approx 6.35$ from $332 ms$ to $52 ms$. Increasing the number of samples to $64$ even leads to a reduction factor of $\approx 8.8$ in comparison to the baseline, which highlights the potential of the hybrid HPC2Edge workflow. \\

\begin{table}[tb]
\centering
\caption{Results at different scales (averaged over five seeds).}
\begin{tabular}{c|c|c|c|c|c}
\textbf{Num. Samples} & \textbf{Val. Score} & \textbf{Val. Loss} & \textbf{Inference Time} & \textbf{Test Score} &\textbf{Test Loss} \\
\hline
0 (baseline)        & 412.81 & 0.0807          & $332.11ms$ &  457.51 &   0.1254 \\
16                  & 146.02 & 0.0937            & $52.30ms$ & 156.66 & 0.1044 \\
32                  & 140.26 & 0.0959             & $44.34ms$ & 140.53 & 0.0962 \\
64                  & 129.99 & 0.0923            & $37.72ms$ & 130.58 & 0.0929      
\end{tabular}
\label{tab:results_table}
\end{table}

\begin{figure}
    \centering
    \includegraphics[width=0.4\linewidth]{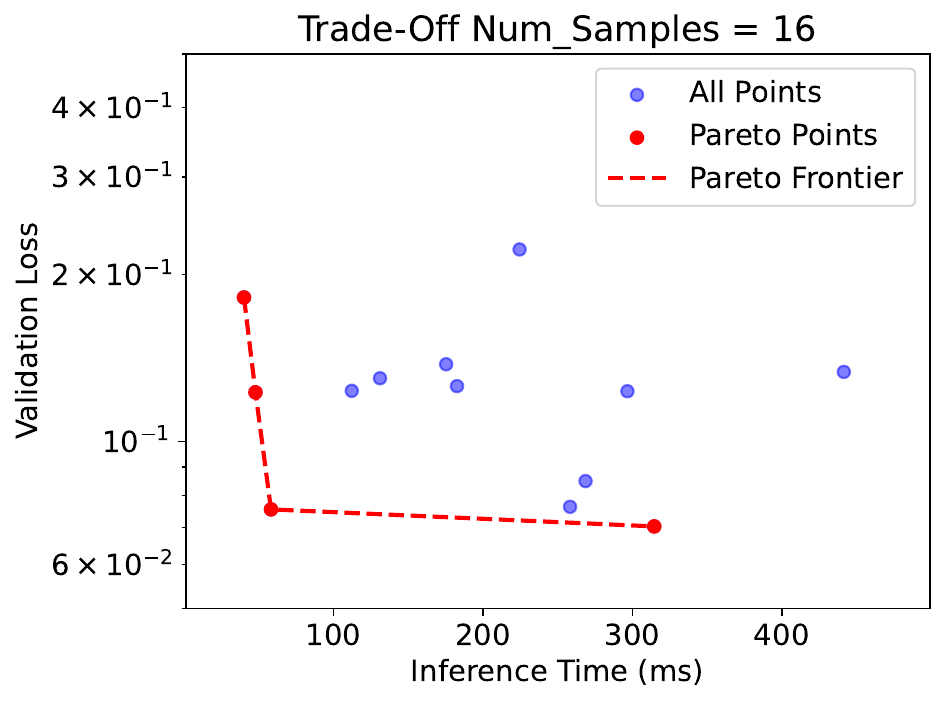}
    \includegraphics[width=0.4\linewidth]{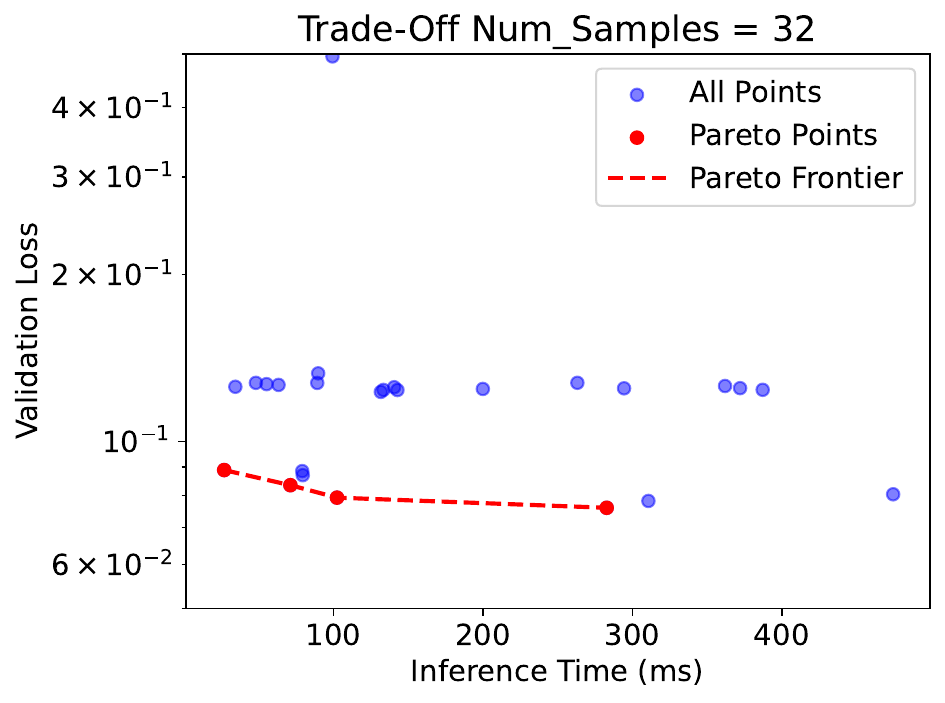}
    \includegraphics[width=0.4\linewidth]{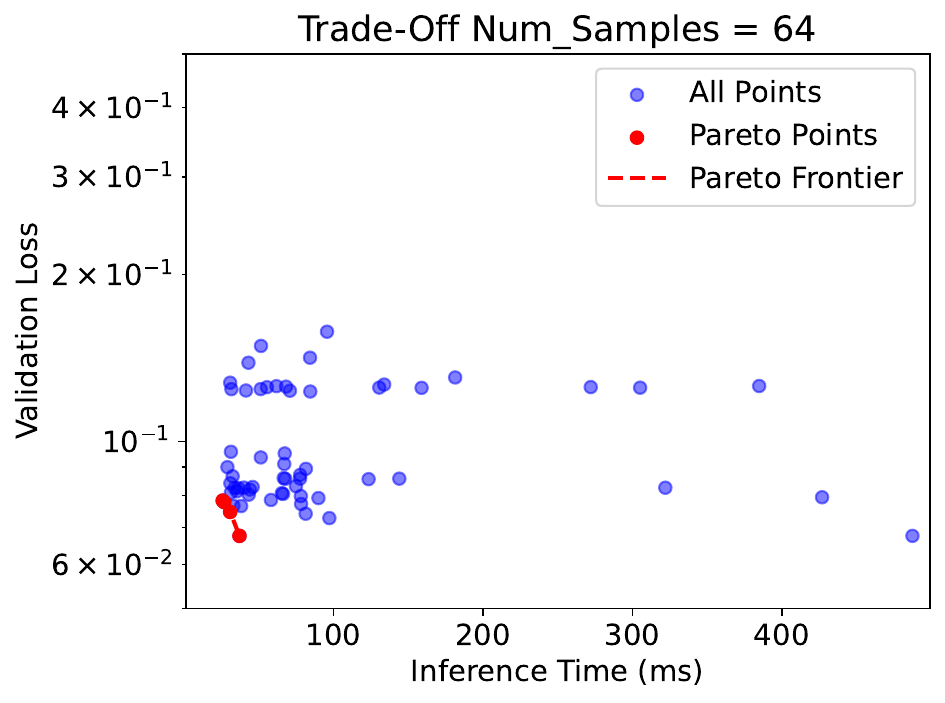}
    \caption{Results at different scales, showing the median best run.}
    \label{fig:pareto_plots}
\end{figure}

In terms of model quality, the validation loss metric shows varying performance across different sample sizes. At $16$ samples, the validation loss increases to $0.0937$, while at 64 samples it still remains higher at $0.0923$, compared to the baseline. In contrast, the $64$ samples \ac{NAS} run results in the best model, decreasing the test loss by a factor of $\approx 1.35$ compared to the baseline model. It is hypothesized that the fluctuations in the validation loss metric are due to the weighting of both metrics into a single one, see Eq.~\eqref{eq:solution_combination}, and thus a larger focus is on the inference time. However, as shown by the decreasing test loss, the model still seems to be able to achieve a higher solution quality than the baseline. Both metrics, the model quality on the test set and the inference time in general decrease for the model on the edge as the number of samples (and thus the compute resources spent on the \ac{NAS} loop) is increased. In Fig.~\ref{fig:pareto_plots}, the pareto curves of the different configurations are depicted, showing the optimal trade-off points between validation loss and inference time. As can be seen, in all cases the pareto curves do move to the left bottom of the plots, indicating better models. \\

To assess the importance of the architectural and optimizer-related parameters on the model performance, the top $10\%$ models found during the 5 \ac{NAS} runs with $64$ samples are examined through their median hyperparameter values. The results show that a small learning rate of $\approx 4.7\cdot 10^{-4}$, combined with a large decay factor of $\approx 0.75$ and a moderate adjustment interval of $20$ steps results in the lowest validation score values. From an input data perspective, a median patch size of $[4,4,4]$ suggests larger patches to be more favorable. From an architectural point of view, a median depth of $[1,1,2,1]$, a median attention head number of $[3, 6, 3, 12]$, and a median embedding dimension of $24$ suggests smaller models to be favorable. This is expected as these parameters usually also result in shorter inference times, which is one of the two objectives of the multi-objective optimizer. 

\section{Summary and Conclusion}\label{sec:summary}
In this work, a hybrid, cross-border, hardware-aware \ac{NAS} workflow that runs in parallel on an \ac{HPC} system and an edge device was presented. The workflow leverages the powerful \acp{GPU} of the \ac{HPC} system to train different model configurations with data-parallel training while performing the inference time measurement of the models on the actual target device, resulting in an accurate measurement. To hide communication latency, each candidate model architecture is sent to the inference device before training on the \ac{HPC} system. The empirical results clearly highlight how large savings in terms of inference time and model quality (in terms of final test loss) can be achieved. As a key finding, this research demonstrates empirically that increasing the computational resources of the \ac{HPO} loop can lead to a smaller computational resource usage during the inference on the edge device. In the future, such workflows could therefore be used to find architectures with even higher speed (compared to expert baselines) or fitting on even smaller edge devices, which is an important feature not only in \ac{AM} but in any field where fast \ac{ML} models are deployed on small devices.

\subsubsection{Acknowledgements.}
The CoE RAISE project has received funding from the European Union’s Horizon 2020 Research and Innovation Framework Programme H2020-INFRAEDI-2019-1 under grant agreement no.~951733. This research was also partially supported by the RELAI project of Flanders Make, the strategic research centre for the manufacturing industry, and by the Flemish Government AI Research Program. Resources for the database were provided by the VSC (Flemish Supercomputer Center), funded by the Research Foundation - Flanders (FWO) and the Flemish Government.

\subsubsection{Disclosure of Interests.}
The authors have no competing interests to declare that are
relevant to the content of this article.

\subsubsection{Venue and Manuscript Version.} This work was accepted for publication in the proceedings of ISC 2025 workshop Computational Aspects of Deep Learning (CADL 2025), and was selected for oral presentation.  This preprint version of the manuscript, however, has not undergone peer review or any post-submission improvements or corrections. The Version of Record of this contribution will be published in HIGH PERFORMANCE COMPUTING: ISC High Performance 2025 International Workshops (Lecture Notes in Computer Science --- LNCS).

\bibliographystyle{splncs04}
\bibliography{hpc2edge}
\end{document}